\documentclass[aps,prb,twocolumn,superscriptaddress,showpacs]{revtex4}
\usepackage[dvips]{epsfig}
\usepackage[dvips]{color}
\usepackage{amssymb}
\usepackage{bm}
\newcommand{\PbNi}{PbNi$_2$V$_2$O$_8$}

\begin{document}

\title{Triplet spin resonance of the Haldane compound with interchain coupling}

\author{A.~I.~Smirnov}
\author{V.~N.~Glazkov}
\affiliation{P.~L.~Kapitza Institute for Physical Problems RAS, 119334
Moscow, Russia}

\author{T.~Kashiwagi}
\author{S.~Kimura}
\author{M.~Hagiwara}
\affiliation{Center for Quantum Science and Technology under Extreme
Conditions (KYOKUGEN), Osaka University, 1-3 Machikaneyama, Toyonaka,
Osaka 560-8531, Japan}

\author{K.~Kindo}
\affiliation{Instiute for Solid State Physics (ISSP), University of Tokyo,
5-1-5 Kashiwanoha, Kashiwa, Chiba 277-8581, Japan}

\author{A.~Ya.~Shapiro}
\author{L.~N.~Demianets}
\affiliation{A.~V.~Shubnikov Institute for Crystallography RAS, 117333
Moscow, Russia}

\date{\today}

\begin{abstract}
   Spin resonance absorption of the triplet excitations is
 studied experimentally in the Haldane magnet \PbNi{}. The spectrum has
features of spin $S=1$ resonance in a crystal field, with all three
components, corresponding to  transitions $|\pm1\rangle \leftrightarrows
|0\rangle$
 and
$|-1\rangle \leftrightarrows |1\rangle$, being observable. The resonance
field is temperature dependent, indicating the renormalization of
excitation spectrum in interaction between the triplets. Magnetic
resonance frequencies and critical fields of the magnetization curve are
consistent with  a boson version of the macroscopic field
theory,\cite{Affleck,FarMar} implying the field induced ordering at the
critical field, while contradict the previously used approach of
noninteracting spin chains.

\end{abstract}

\maketitle

\section{Introduction}

The inorganic dielectric \PbNi{}, with the magnetic structure formed by
chains of Ni$^{2+}$ ($S$=1) ions, exhibits a Haldane like ground
state.\cite{Uchiyama,Zheludev1} Due to anisotropy and interchain exchange
the Haldane energy gap is reduced and \PbNi{} is close to the critical
point of the quantum phase transition from a spin-liquid to an ordered
easy-axis antiferromagnet.\cite{Sakai} Besides, the spin-liquid phase may
become unstable at the critical magnetic field, corresponding to the
vanishing of the energy gap of triplet excitations. In Heisenberg exchange
approximation the critical field value $H_c$ and the energy gap $\Delta$
are related by a simple relation $g\mu_B \mu_0 H_c=\Delta$. The influence
of the single-ion anisotropy on the Haldane spin chains was analysed
theoretically by the exact diagonalization for finite
chains,\cite{Golinelli} by a perturbative
approach,\cite{Golinelli,Zaliznyak} as well as by macroscopic field theory
methods.\cite{Affleck} In the magnetic field range far below $H_c$ all
models result in the same energy levels, parametrized  by the main gap,
one or two anisotropy constants and $g$-factor. The perturbative
description of excitations in the Haldane chain appears to be identical to
the description of an isolated spin $S=1$ in a crystal
field.\cite{Abraham} Extrapolated to the field of spin-gap closing, the
perturbative approach yields critical fields
\begin{equation}
g \mu_B \mu_0 H_{\alpha c} = \sqrt{\Delta_{\beta}\Delta_{\gamma}},
\label{formula:Hc}
\end{equation}
\noindent here the magnetic field is applied along principal direction
$\alpha$, while $\beta$ and $\gamma$ note other principal directions.
 For the uniaxial case the $S_z=\pm 1$ triplet
components with the momentum of $\pi$ have equal gaps
$\Delta_{x}=\Delta_{y}$, and  the gap of $S_z=0$ component, $\Delta_{z}$,
is higher or lower depending on the anisotropy type.
 The value of the gap splitting caused by the crystal
field is, in terms of $S=1$ problem,  the effective anisotropy constant
$D_{eff}= \Delta_{x}-\Delta_{z}$.  For Haldane chains $D_{eff} = -1.98 D$
and $(2\Delta_{x}+\Delta_{z})/3)=\Delta$, where $D$ is a single ion
anisotropy constant contributing to the single ion Hamiltonian with the
term $DS_z^2$.

The spectrum of a real spin-gap system may be more complicated, as, e.g.,
for the Haldane magnet Ni(C$_{2}$H$_{8}$N$_{2}$)$_{2}$NO$_{2}$(ClO$_{4}$)
(abbreviated as NENP), when the spin gap remains non-zero due to the
staggered $g$-tensor,\cite{NENPSieling} or for the dimer spin-gap system
TlCuCl$_3$ \cite{Kolezhuk,FarMar} with a nonlinear frequency-field
dependence near the critical field, at which the field-induced 3D ordering
occurs, {\it etc}. \PbNi~ is a convenient model system for study the
influence of 3D correlations and anisotropy on the spin-gap in a magnetic
field, because this compound demonstrates a spin-gap split by the
anisotropy, spin-liquid behavior below the critical field and
field-induced 3D antiferromagnetic ordering above $H_c$.\cite{Tsuji}

The perturbative approach of noninteracting Haldane chains (in particular,
formula (\ref{formula:Hc})) was used \cite{Zheludev2} for \PbNi~ to derive
the energy gaps from the critical field values. Further, the velocity of
spin excitations, intrachain and interchain exchange integrals were found
by fitting the inelastic neutron scattering intensity from a powder sample
keeping the above spin gap values fixed.
\cite{Uchiyama,Zheludev1,Zheludev2} Nevertheless, the extrapolation of the
perturbative approach to high fields may be doubtful. While it is
justified by the exact diagonalization of the finite chain problem,
\cite{Golinelli} the field-theory treatment \cite{Affleck} has resulted in
two models, diverging in high fields. The boson model should be exact in
presence of interchain coupling of nearly a critical value, while fermion
model is exact for a hypothetic Hamiltonian. Analogous theory considering
a general case of a spin-gap magnet with the field-induced 3D ordering was
developed in Ref.\onlinecite{FarMar}. The source Lagrangian and relations
between the critical fields in the bosonic model\cite{Affleck} and in the
macroscopic theory\cite{FarMar} are identical. The results of the
fermionic model are close to the perturbative theory and exact
diagonalization of finite chain problem, while the boson model and the
model \cite{FarMar} diverge from the perturbative curves near the critical
field. Probably, this divergency should be ascribed to fluctuations, which
are suggested to be suppressed (e.g. by 3D coupling) in the boson model of
Ref. \onlinecite{Affleck} and in the macroscopic theory  of Ref.
\onlinecite{FarMar}.

In the present paper we study the triplet excitations by means of the
high-frequency electron spin resonance (ESR) spectroscopy which enables
one to measure directly the frequencies of the transitions between the
spin sublevels of the collective triplet states. We recover the splitting
of the zero-field energy gap from the experiment and relate it to the
difference between the critical fields $H_{xc}$ and $H_{zc}$. The
zero-field splitting of the triplet levels appears to be about one-half of
the value estimated by the perturbative relation(\ref{formula:Hc}). At the
same time, the spectrum of triplet excitations is well described by
macroscopic models, implying 3D ordering at the critical
field.\cite{Affleck,FarMar} Thus, the influence of the interchain coupling
was found to be of importance for the magnetic excitation spectrum in the
vicinity of the critical field.

\section{Experiment}

\begin{figure}
\centering \epsfig{file=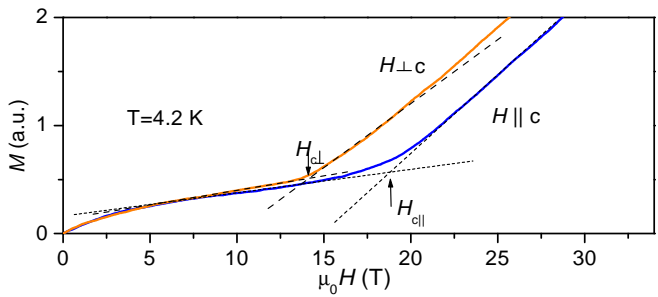, width=1.0\columnwidth, clip=}
\caption{Pulse field magnetization of the aligned sample of \PbNi{}. Solid
curves: experiment, dashed stright lines - extrapolations to
$H_{c\perp,\parallel}$ } \label{fig:MvsH}
\end{figure}

We used ceramic samples of  PbNiV$_2$O$_8$ prepared as described in
Ref.\onlinecite{PRBSmirnov2002} and magnetic field aligned samples
prepared following Ref. \onlinecite{Uchiyama} with the orientation of
$c$-axis of crystallites  parallel to the aligning field. Magnetization
curves taken by the pulse technique are shown in Fig.~\ref{fig:MvsH}. The
critical field values derived from these curves are $\mu_0
H_{c\parallel}$=19 $\pm 0.5$ T  and $\mu_0 H_{c\perp}$=14$\pm 0.5$ T,
which are well consistent with the data of Refs.\onlinecite{Uchiyama} and
\onlinecite{Tsuji}. The ESR spectra were measured both for ceramic and
aligned samples, as microwave transmission {\it vs} magnetic field
dependences. Magnetic field was created by a 16 T cryomagnet. Microwave
signal in the frequency range 70-500 GHz was generated and detected by
means of the vector network analyzer from the ABmm system at the KYOKUGEN
Center.

\begin{figure}
\centering \epsfig{file=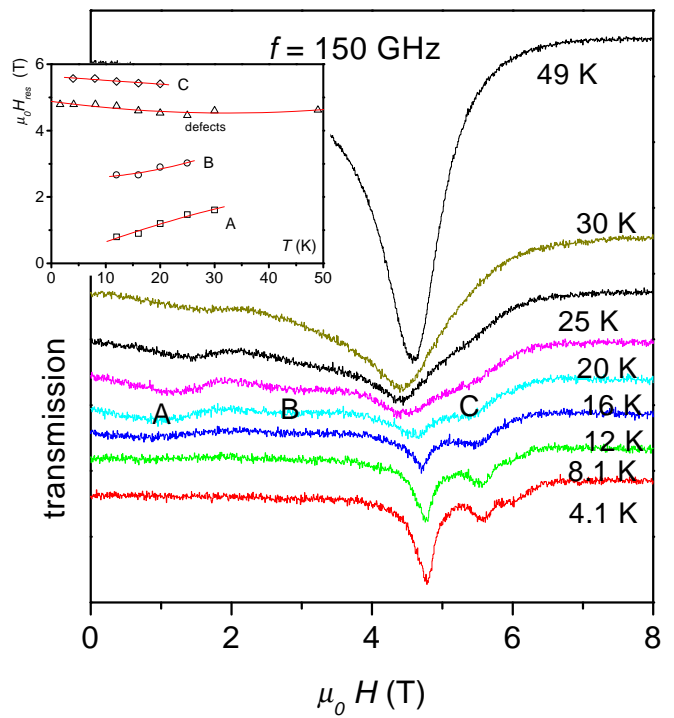, width=1.0\columnwidth, clip=}
\caption{150 GHz ESR spectra of a \PbNi{} ceramic sample measured at
different temperatures. Inset: Temperature dependence of the ESR fields A,
B, C.  Lines are guide to the eye. } \label{fig:150GHzvarT}
\end{figure}

\begin{figure}
\centering \epsfig{file=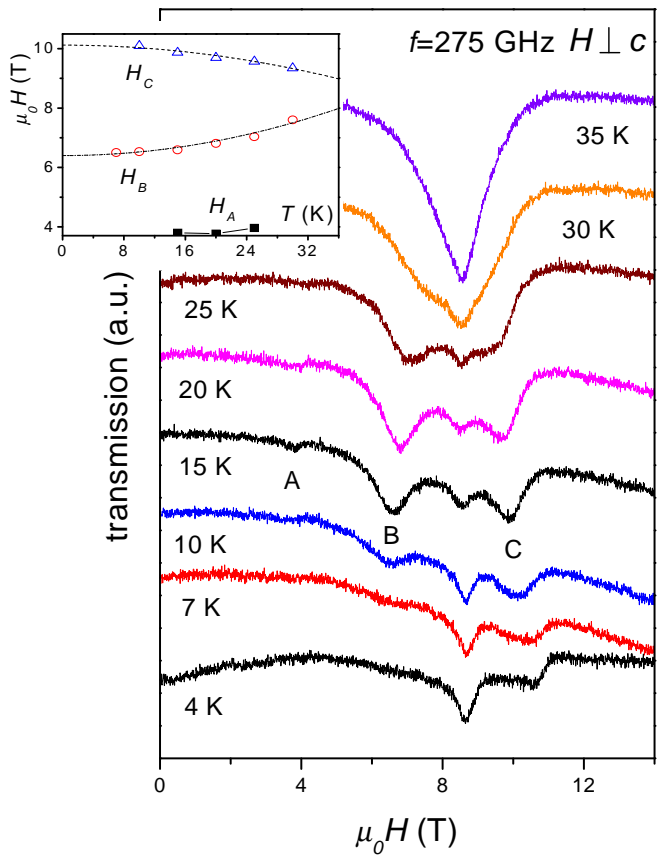, width=1.0\columnwidth, clip=} \caption{
275 GHz ESR spectra of a \PbNi{} aligned sample measured at different
temperatures for {\bf \it{H}} perpendicular to the alignment axis. Inset:
Temperature dependence of the ESR fields A, B, C. Lines are guide to the
eye. } \label{fig:275GHzvarT}
\end{figure}

A typical ESR spectrum of a ceramic sample and its evolution with
temperature is presented for the frequency $f=$150 GHz on  Fig.
\ref{fig:150GHzvarT}.~ Analogous spectra for the aligned sample and the
frequency of 275 GHz are presented on Fig.~\ref{fig:275GHzvarT}.  At the
temperatures above 30 K there is a single ESR line, corresponding to a
typical exchange-narrowed Ni$^{2+}$ ESR signal with $g$-factor value of
2.23. At cooling the sample we observe the diminishing of the intensity of
this signal and appearing of three ESR lines apart from the high
temperature ESR field. These signals are marked by letters A, B and C on
Figs.~\ref{fig:150GHzvarT} and \ref{fig:275GHzvarT}. They disappear again
at low temperatures, demonstrating a temperature dependent shift of the
resonance field, as shown on the Insets of Figures. At low temperatures an
upturn of the resonance absorption in g-factor range 1.9-2.2 with the
maximum at at $g=2.2$ occurs. This low-temperature signal is due to
defects and impurities. The intensity of marked signals, increasing with
temperature, indicates a thermally activated nature of this kind of
resonance.  The separated lines of thermally activated absorption are
distinctly seen in the temperature range 15-20 K, at lower temperatures
the intensity is weak because of the spin gap, and at higher temperatures
the lines merge into a single line. No absorption signal corresponding to
the singlet-triplet transitions is observed.

\begin{figure}
\centering \epsfig{file=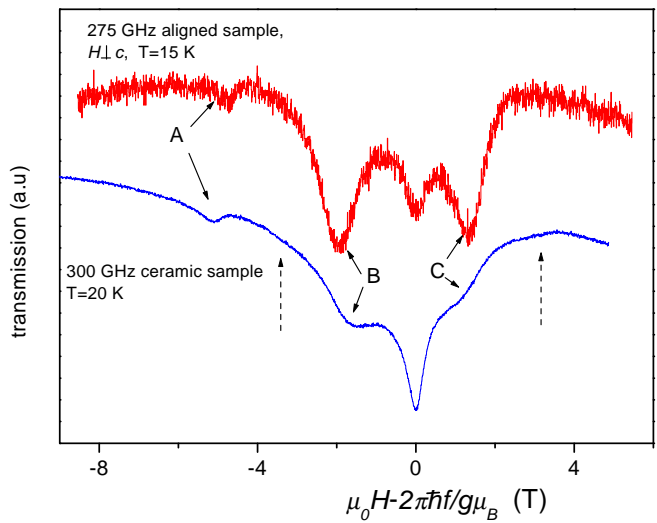, width=1.0\columnwidth, clip=} \caption{
300 GHz ESR spectrum of a \PbNi{} ceramic sample at $T=$20~K (lower curve)
and 275 GHZ spectrum of an aligned sample (upper curve).}
\label{fig:300&276GHz}
\end{figure}

\begin{figure}
\centering \epsfig{file=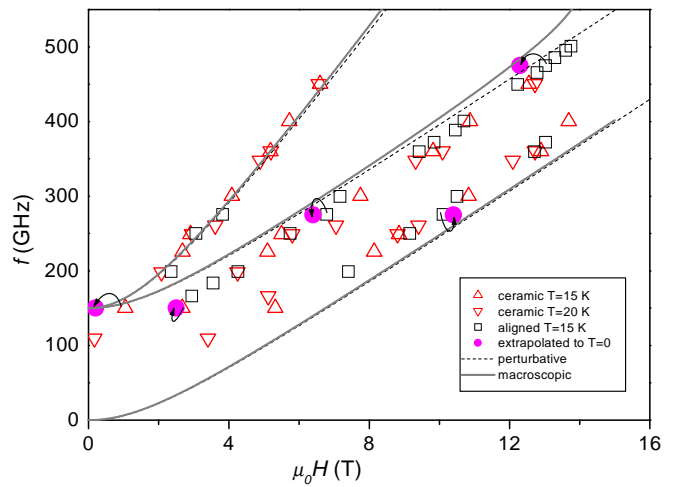, width=1.0\columnwidth, clip=} \caption{
Frequency-field diagram of the thermally activated magnetic resonance of
\PbNi{} ceramic and aligned samples. Triangles and squares present
experimental data for resonances marked by A, B, C. Lines are model
calculations. Solid circles represent the extrapolations to $T$=0 noted by
curved arrows.} \label{fig:fvsHnew}
\end{figure}

Comparison of the ESR absorption of the ceramic and aligned samples at
close frequencies is given on Fig. \ref{fig:300&276GHz}.  Dashed arrows
near the curve, corresponding to  the ceramic sample, mark boundaries of
the absorption band and letters A, B, C denote thermally activated lines
for both samples. The central peak is ascribed to defects and diminishes
with temperature as a conventional paramagnetic resonance line. The
analogous record of the aligned sample at $\bm{H} \perp c$ has more
distinct resonance lines at the positions B and C, than the ceramic
sample. The positions of the local maxima of thermally activated
absorption, taken at different frequencies, are plotted on Fig.
\ref{fig:fvsHnew}.

\section{Discussion}
\subsection{General aspects}

The lineshape of the thermally activated absorption (Figs.
\ref{fig:150GHzvarT} and \ref{fig:300&276GHz}) corresponds approximately
to the known ESR spectrum of a powder and crystal samples with $S=1$ ions
in an axial crystal field (see, e.g., Ref. \onlinecite{Abraham}). For a
uniaxial crystal, this ESR spectrum, in case of $g\mu_B \mu_0 H \gg D$,
consists of three resonance lines corresponding to  the transitions
$\mid\pm1\rangle\leftrightarrows\mid 0\rangle$
 with  $\Delta S_z =\pm 1$ and
$\mid -1\rangle\leftrightarrows\mid 1\rangle$ with $\Delta S_z=\pm 2$. Two
 resonance lines, corresponding to $\Delta S_z =\pm 1$, have larger
intensity and are on both sides from the free spins resonance field $\mu_0
H_0=2\pi\hbar f/(g\mu_{B})$. The distance between these two lines is
 $2D/g\mu_B$ for $\bm{H}
\parallel c$ and approximately $D/g\mu_B$ for $\bm{H} \perp c$.
 Here $D$ is a single ion
anisotropy constant contributing to the single ion Hamiltonian with the
term $D\hat{S}_z^2$. Naturally, for a powder sample there should be a band
of absorption instead of resonance lines, the whole band width is
$2D/g\mu_B$, and the maxima of absorption are near the resonance field of
crystallites with $c \perp \bm {H}$ because of their largest statistical
weight (the field interval between maxima is again $D/g\mu_B$). In case of
the easy plane anisotropy the intensity of the lower field maximum is
larger than that at the upper one because of a higher population of the
lower spin sublevels. For the case of the easy axis anisotropy the
relation between the intensities should be inversed.

 Applying the above consideration of $S=1$ magnetic ion in a crystal
field we conclude, that the effective anisotropy of the triplet
excitations is of the easy plane type ($D_{eff}
> 0$). According to Ref.~\onlinecite{Golinelli}, $D_{eff}$ and $D$ are
of different signes, thus, the single-ion constant $D$ should be negative,
which agrees with the easy axis anisotropy of the impurity induced ordered
phase. \cite{PRBSmirnov2002}

Maximums of the absorption are separated by approximately 3.5 T, which
corresponds to $D_{eff}\simeq 110$~GHz (here and farther on we assume
$g=2.23$, as proved by ESR at $T>30$K). The leftmost absorption line
demonstrates a characteristic frequency-field dependence (Fig.
\ref{fig:fvsHnew}) with asymptote $f=2g\mu_B \mu_0 H/(2\pi\hbar)$, marking
a two-quantum transition with $\Delta S_z = \pm2$. Its zero-field gap is
equal to the splitting of the triplet sublevels\cite{Abraham}. Observation
of this absorption at the frequency 150 GHz indicates directly, that the
splitting of the triplet levels by the effective crystal field is less
than or about 150GHz.

\begin{figure}
\centering \epsfig{file=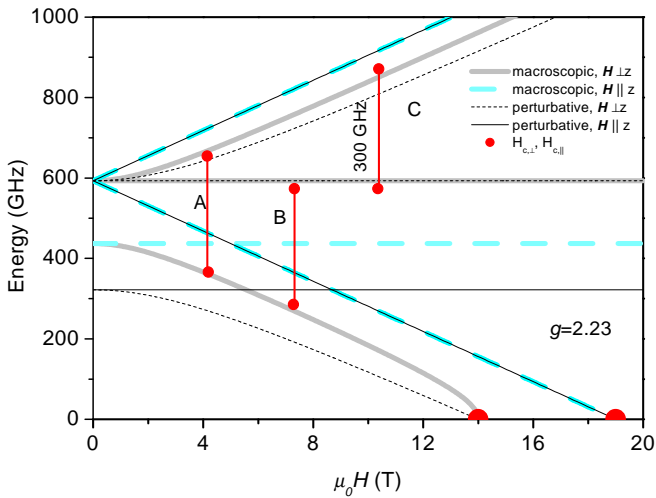, width=1.0\columnwidth, clip=} \caption{
Triplet energy levels in a spin gap magnet with $\mu_0 H_{c\perp}$=14.0 T
and $\mu_0 H_{c\parallel}$=19.0 T, calculated for noninteracting Haldane
chains\cite{Golinelli} (thin lines) and derived in the macroscopic model
\cite{FarMar} (thick lines). Vertical segments present observed
transitions at $f$=300 GHz} \label{fig:levnew}
\end{figure}

\begin{figure}
\centering \epsfig{file=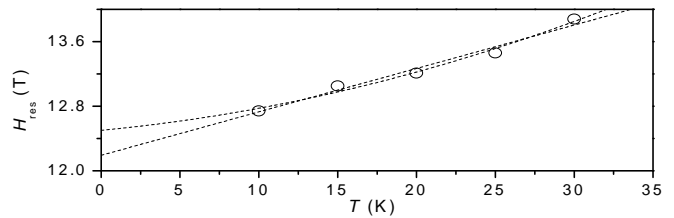, width=1.0\columnwidth, clip=}
\caption{Temperature dependence of the resonance field of the mode B at
the frequency 475 GHz for the aligned sample at ${\bf H} \perp c$. Dashed
lines are linear and second order polynom extrapolations to zero
temperature. } \label{fig:extrapol}
\end{figure}

The perturbative approach \cite{Golinelli}  yields by (\ref{formula:Hc})
the zero-field gaps $E_{min1}=320\pm10$~GHz and $E_{min2}=590\pm10$~GHz.
The corresponding splitting of 270~GHz is twice as large as estimated from
the observed ESR lines. Alternatively, energy levels of the triplet
excitations can be calculated within the boson model,\cite{Affleck,FarMar}
which imply 3D ordering above the critical field. Details of the
calculation are given in Appendix. For $D_{eff}>0$ the gaps are related to
the critical fields as follows:

\begin{equation}
 g\mu_B \mu_0 H_{c \perp}=E_{min1},~~~~~~~~~g\mu_B  \mu_0 H_{c\parallel}=E_{min2}\\
 \label{eqn:HcFarMar}
\end{equation}

From the critical field values we get here $E_{min1}=440\pm10$~GHz and
$E_{min2}=590\pm10$~GHz. The corresponding splitting of  150~GHz is in
much better agreement with the above experimental results. The calculated
triplet sublevels for perturbative and macroscopic models are presented at
the Fig.\ref{fig:levnew}. Here all theories are locked to the observed
values of critical fields. The difference in  $E_{min1}$ occurs due to the
nonlinear field dependence of the energy of the lowest triplet sublevel
near $H_{c\perp}$ in the macroscopic models. It should be noted, that in
low fields, i.e. far from the critical point, all models would demonstrate
identical results when they would locked on the same energy gaps at $H=0$.

The frequency-field dependence for the case $\bm{H}\perp c$ calculated in
a perturbative approach and in the macroscopic models is presented in Fig.
\ref{fig:fvsHnew}. It should be noted that no fitting parameters are used
here, the value of zero field splitting, $D_{eff}=E_{min2}-E_{min1}$ is
taken from critical fields using relations (\ref{eqn:HcFarMar}).The
resonance fields were taken at  $T=$15 and 20~K, while critical fields are
measured at 4.2 K. Thus, we have to extrapolate the observed resonance
fields to zero temperature. The temperature evolution of the ESR spectra,
analogous to that shown in Fig.~\ref{fig:275GHzvarT}, was followed for the
frequencies 150, 275 and 475 GHz. The resonance fields, extrapolated to
$T$=0 are shown on Fig.\ref{fig:fvsHnew} by solid circles, pointed by
curved arrows. The data, extrapolated to zero temperature, lie on the
model curve within the experimental error.

Note that the energy levels predicted in Refs. \onlinecite{Uchiyama},
\onlinecite{Zheludev1} and \onlinecite{Zheludev2} by use of the
perturbative model\cite{Golinelli} ($D_{eff}=270$GHz and
$E_{min2}/(2\pi\hbar)=$590 GHz) does not correspond the above experimental
data to any extent. Thus, the experimental ESR frequencies are in a
agreement with the energy levels predicted by the macroscopic theories,
\cite{Affleck,FarMar}, rather than with the theory considering isolated
Haldane chains.\cite{Golinelli}

\subsection{Modeling}

\begin{figure}
\centering \epsfig{file=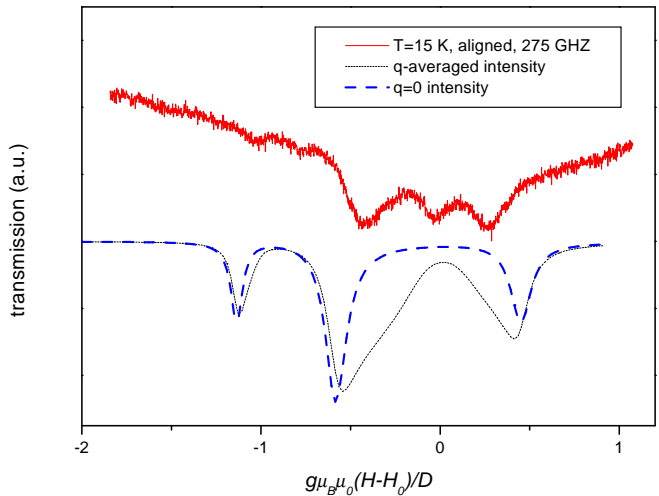, width=1.0\columnwidth, clip=}
\caption{275 GHz ESR spectrum of a \PbNi{} aligned sample at $T=$15~K,
$\bm{H} \perp c $, and the calculated spectrum in $S=1$ model for the
triplets at the bottom of the excitations zone (q=0) and a spectrum,
averaged over the excitations, excited at $T$=20 K. Parameters used are:
$E_{min2}$=590 GHz, $D_{eff}(0)$= 150 GHz, ESR linewidth 10 GHz, $s$=4895
GHz, $g=2.23$ } \label{fig:model}
\end{figure}

\begin{figure}
\centering \epsfig{file=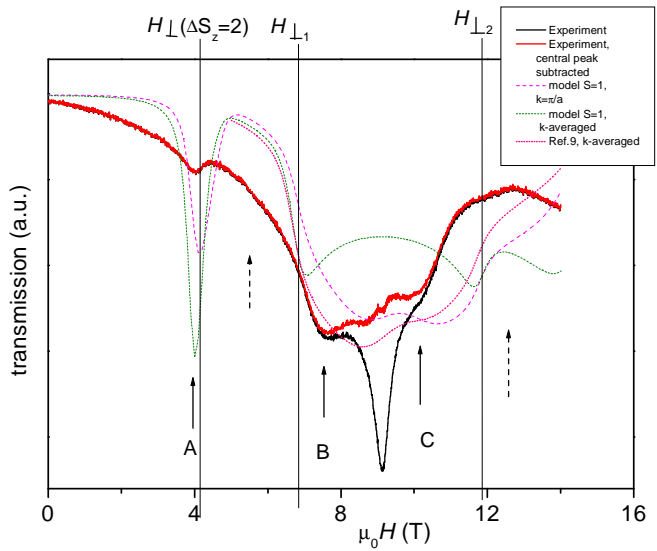, width=1.0\columnwidth, clip=}
\caption{300 GHz ESR spectrum of a \PbNi{} ceramic sample at $T=$20~K
(black solid line), the same spectrum with the defects Lorentzian line
subtracted and model spectra as described in the text. Vertical lines are
at the resonance fields  for crystallites with the field lying in the
basal plane. Parameters used are: $E_{min2}$=590 GHz, $D_{eff}(0)$= 150
GHz, ESR linewidth 10 GHz, $s$=4895 GHz, $g=2.23$ } \label{fig:300GHz}
\end{figure}

We performed a model calculation of the ESR absorption for the powder
sample of a spin gap magnet with gapped triplet excitations (see Figs.
\ref{fig:300GHz} and \ref{fig:model}). The modeling was performed by two
methods, giving practically similar results for the lineshape and
positions of the maxima of ESR absorption.

For the first method we considered the energy levels of effective spin
$S=1$ ascribed to a  triplet excitation mode with a wavevector $\bm {k}$.
The population numbers for the $S_z=\pm1$ and $S_z=0$ spin sublevels of a
triplet excitation mode were calculated  according to the Plank's
distribution function. The effective anisotropy constant, equal to the
zero field splitting of a triplet  sublevels is defined as

\begin{equation}\label{eqn:Deff(k)}
  D_{eff}(\bm{k}) = \hbar
  (\omega_{x}(0,\bm{k})-\omega_{z}(0,\bm{k}))
\end{equation}

\noindent here $\omega_{x,y}(0,\bm {k})$ and $\omega_{z}(0,\bm{k})$ are
zero-field cyclic frequencies of the triplet excitations carrying
effective spin projection $S^{eff}_z$ equal to $\pm 1$ or zero
correspondingly. We calculated the microwave absorption for effective
spins $S=1$ in a uniaxial crystal field according the standard ESR
procedure, i.e. calculating transition frequencies, matrix elements of the
transitions and difference of the population numbers \cite{Abraham}. The
spectrum of the triplet excitations in zero field  is taken in the form
\cite{Zheludev2,Affleck,FarMar}:

\begin{eqnarray}
  \omega_{x,y}(\bm{q})&=& \sqrt{\omega^2_{x,y}(0)+s^2 q^2}  \\
  \omega_{z}(\bm{q}) &=& \sqrt{\omega^2_{z}(0)+s^2 q^2}
\end{eqnarray}

with $q=\pi-ka$,  $s$=4895 GHz (Ref.\onlinecite{Zheludev1}) , and
$\omega_{z}(0)/(2\pi)$= 440 GHz, $\omega_{x,y}(0)/(2\pi)$=590 GHz derived
from the fields $H_{c\parallel}$ and $H_{c\perp}$ according to Ref.
\onlinecite{FarMar} as shown on Fig. \ref{fig:levnew}. The field
dependences of $\hbar\omega_{\alpha}(k)$  are derived as the energies of
corresponding  sublevels of an effective spin $S=1$ in a crystal field
with the anisotropy constant $D_{eff}(k)$.

This modeling also provides the low-field $\Delta S=\pm 2$ transition,
which is allowed in the canted orientations of the magnetic field with
respect to anisotropy axis.

For the second method we used the excitations frequencies, given by the
macroscopic models.\cite{Affleck,FarMar} These frequencies can be
calculated for an arbitrary $\bm{H}$ and $\bm{q}$ (see Appendix) in the
low frequency approximation $\hbar\omega_{\alpha} \ll J$. Since this model
does not provide information on matrix elements of the transitions, which
are necessary to calculate intensity of the corresponding ESR absorption,
we have used this model only at high fields taking the matrix elements for
$\Delta S_z=\pm1$ transitions to be equal and field independent.
Population numbers of the exited states were taken according to the
Plank's distribution function again.

The model microwave absorption due to the excitations with the energy at
the bottom of the zone (q=0)  is shown in Fig. \ref{fig:model} along with
the model absorption curve  at T=15~K, when the excitations away of the
zone bottom are also excited. Despite of the smaller value of $D_{eff}$ of
the high energy excitations, the peaks of absorption remain at the same
positions due to a large spectral density at the zone bottom. At heating,
the absorption curve becomes asymmetric with enlarging of the intensity in
the field range, corresponding to resonance for smaller $D_{eff}$, as
predicted in Ref. \onlinecite{Affleck}.  For modeling of the ESR
absorption of the ceramic samples, we made the  averaging over the random
orientations of crystallites  and, again, over the one-dimensional
$\mathbf{k}$-space.  The calculated lineshape, presented by the dashed
curve on Fig. \ref{fig:300GHz} corresponds qualitatively to the observed
line, if we subtract the central peak of absorption, produced by the
defects, as shawn on the Fig. It should be noted that the singularity in
the form of the step of absorption is close to the maximum of absorption
for the single-$\bm{k}$ calculation, both singularities are due to the
maximum of the statistic weight of the crystallites with the four fold
axis perpendicular to the magnetic field. For the $\bm{k}$-space
integrated absorption the steps are at the same positions, while the
maxima are smeared due to the contribution of the higher energy
excitations which have smaller value of $D_{eff}(\bm{k})$. The results of
the modeling are practically independent on the natural ESR linewidth
changing in the range 1--10 GHz.

\subsection{Temperature dependent ESR}
The temperature dependence of the triplet ESR field observed in the
present work  may be, probably, attributed to an interaction between the
triplets, analogous to the spin-wave frequencies in conventional ferro-and
antiferromagnets, which is known to be renormalized with excitation of
magnons (see, e.g., Refs. \onlinecite{Dyson} and
\onlinecite{ProzorovaSmirnov}). The temperature dependence of the triplet
excitation spectrum  was observed also for the dimer spin-gap compound
TlCuCl$_3$.\cite{GlazkovPRB} An alternative reason for the temperature
dependence of $D_{eff}$ and the triplet excitation spectrum may be the
change of the correlation length with heating. However, according to
experiments \cite{corrlength}, the correlation length is reduced only for
20 \% at the temperature of a half of the intrinsic Haldane gap $0.41 J$,
as in our case. Apparently, this may not substantially change the
effective anisotropy constant of triplet excitations.  The excitation of
the triplets  apart from the bottom of the excitation spectrum also may
not change the position of the absorption maxima, despite they should have
smaller $D_{eff}$. As noted in Ref. \onlinecite{Affleck}, the influence of
higher energy excitations will result in an asymmetric form of the ESR
lines but not in the shift of the absorption maximum.

\section{Conclusions}

 In conclusion, the ESR spectrum  of triplet excitations in a Haldane magnet
 \PbNi~demonstrates a temperature dependence, probably, due to the interaction
 between the excitations.  ESR of triplet excitations
 is  satisfactory modeled without fitting parameters, using the
experimental  values of critical fields. The relation between the critical
fields and energy gaps corresponds to the boson model of
Ref.~\onlinecite{Affleck} and macroscopic theory\cite{FarMar} and
contradicts the fermion model of Ref.~\onlinecite{Affleck} and isolated
chain calculations of Ref.~\onlinecite{Golinelli}. This indicates,
supposedly, a strong influence of the 3D correlations on the spectrum of
excitations in \PbNi.

\acknowledgments

Authors are indebted to  A.~M.~Farutin, V.~I.~Marchenko,  T.~Sakai and
M.~E.~Zhitomirsky for discussions. This work was in part supported by the
Grant-in-Aid for Scientific Research on Priority Areas from the Japanese
Ministry of Education, Culture, Sports, Science and Technology and  by the
Russian Foundation for Basic Research grant 06-02-16509. Some of these
studies were done under a Foreign Visiting Professor Program in KYOKUGEN,
Osaka University.

\section{Appendix}

Following Ref.\onlinecite{FarMar}, the Lagrangian for the vector field
${\bm \eta}$ corresponding to $S=1$ excitation  for the crystal with the
uniaxial symmetry is

\begin{equation}
L=\frac{1}{2}(\dot{{\bm \eta}}+\gamma[{\bm \eta}{\bm H}])^2
-\frac{A}{2}{\bm \eta}^2- \frac{G_{ij}}{2}\partial_i{\bm \eta}
\partial_j{\bm \eta}+\frac{b}{2}(\eta_x^2 + \eta_y^2 -
2\eta_x^2),
\end{equation}

here $\gamma =g \mu_B \mu_0/\hbar$.

Corresponding equation of motion has the form

\begin{eqnarray}
\ddot{{\bm \eta}} +2 \gamma[\dot{{\bm \eta}}{\bm H}]-\gamma^2H^2{\bm
\eta}+ \gamma^2{\bm H}({\bm \eta}{\bm H})+A{\bm \eta}-  \nonumber \\
 -G_{ij}\partial_i\partial_j{\bm \eta}
-b \left(
    \begin{array}{c}
    \eta_x\\ \eta_y\\ -2\eta_z\\
    \end{array}
    \right)  =0.
\end{eqnarray}

The frequencies at $k=\pi/a$ and $H=0$  are

\begin{eqnarray}
 \omega_{x,y}^{(0)}&=&\sqrt{A_{xx}}=\sqrt{A-b} \\
 \omega_z^{(0)}&=&\sqrt{A_{zz}}=\sqrt{A+2b}
\end{eqnarray}

The field and $k$-dependences for principal directions of the magnetic
field are:

\noindent for ${\bm H}\parallel z$:

\begin{eqnarray}
 \omega_{x,y} &=& \sqrt{A_{xx} + s^2q^2} \pm \gamma H \\
 \omega_{0} &=& \sqrt{A_{zz} + s^2q^2}
\end{eqnarray}

\noindent for ${\bm H}\parallel x$:
\begin{eqnarray}
    \omega^2_{x,y} &=& \gamma^2H^2 +\frac{A_{xx}+A_{zz}}{2} +
    s^2q^2\pm
\nonumber \\
    &&\pm \bigglb[\frac{\left(A_{xx}-A_{zz}\right)^2}{4} +
2\gamma^2H^2\left(A_{xx}+A_{zz}\right) + \nonumber
\\ &&+4\gamma^2H^2 s^2q^2\biggrb]^{1/2}
\\
\omega_{0} &=& \sqrt{A_{zz} + s^2q^2}
\end{eqnarray}

\noindent here $q=k-\pi/a$.

For an arbitrary orientation the frequencies may be found  by solving the
secular equation

\begin{eqnarray}
x^3+(2h^2+A_{zz} - A_{xx})x^2 -h^2\biglb(4(A_{xx}+q^2s^2)- \nonumber
\\
-h^2+(A_{zz}-A_{xx})(1-3\cos^2\theta)\bigrb)x - \nonumber \\
-(A_{zz}-A_{xx})h^2\cos^2\theta \biglb(4(A_{xx}+q^2s^2)-h^2\bigrb)=0
\end{eqnarray}

here $h=\gamma H$ , $x=-\omega^2 +q^2s^2 +A_{xx}$, $\theta $ is the angle
between the magnetic field and $z$-axis.

Critical field in the arbitrary orientation of the applied field for the
effective easy-plane anisotropy of triplet excitations ($b<0$) is given by
the equation

\begin{equation}\label{eqn:Hc(theta)1}
  \gamma H_c(\theta)=\frac{\omega_z^{(0)}\omega_{x,y}^{(0)}}{\sqrt{\left(\omega_{x,y}^{(0)}\right)^2+\left(\left(\omega_{z}^{(0)}\right)^2-\left(\omega_{x,y}^{(0)}\right)^2\right)\cos^2\theta}}
\end{equation}

\noindent while for the effective easy-axis anisotropy of triplet
excitations ($b>0$) $H_c$ is orientation independent

\begin{equation}\label{eqn:Hc(theta)2}
 \gamma H_c=\frac{\hbar}{g\mu_B}\omega_{x,y}^{(0)}
\end{equation}

\end{document}